\documentstyle[twocolumn,epsfig,citesort]{article}

\advance \topmargin by -\headheight
\advance \topmargin by -\headsep
\if@twoside
\else
    \evensidemargin \oddsidemargin
\fi
\DeclareFixedFont{\xsf}{OT1}{cmss}{m}{n}{10}

\begin{document}

\def \lesssim{\leavevmode{\raisebox{-.5ex}{ $\stackrel{>}{\sim}$ } } }
\def \gtrsim{ \leavevmode{\raisebox{-.5ex}{ $\stackrel{>}{\sim}$ } } }
\def \subarr{  \begin{array}{c} \mbox{\scriptsize{$ \{ n_{\ell}\}$ }} \\ 
        \mbox{\scriptsize{ $\{m_{\ell}\}$}}             \\ 
        \mbox{\scriptsize{ $\{p_{\ell}\}$ }}  \end{array} }
\def \subun{ \leavevmode{\raisebox{-1.6ex}{$\stackrel{
	\textstyle{\ell_i,\ell_j}}
	{\mbox{\scriptsize{unconstrained}}} $}} }
\def \maxj{\begin{array}{c} \mbox{max} \\ 
        \mbox{\scriptsize{ $j$}}             \\ 
        \mbox{\scriptsize{ $ 0 \leq j \leq \jmax $ }}  \end{array} }
\def \nl{ \textstyle{\mbox{\scriptsize{$\{n_{\ell}\}$}}} }
\def \ml{ \mbox{\scriptsize{$\{m_{\ell}\}$}} }
\def \pl{ \mbox{\scriptsize{$\{p_{\ell}\}$}} }
\def \be{\begin{equation}}
\def \ee{\end{equation}}
\def \bea{\begin{eqnarray}}
\def \eea{\end{eqnarray}}
\def \zn{z_{\mbox{\tiny N}}}
\def \jmax{j_{\mbox{\tiny MAX}}}
\def \ei{\epsilon_i}
\def \ej{\epsilon_j}
\def \Qj{Q_j}
\def \Qk{Q_k}
\def \Qi{Q_i}
\def \Qu{Q_{\mbox{\tiny U}}}
\def \Qsurf{Q_{\mbox{\tiny SURF}}}
\def \eio{\epsilon_i^{\circ}}
\def \ejo{\epsilon_j^{\circ}}
\def \eabn{\epsilon_{\alpha\beta}^{\mbox{\tiny N}}}
\def \eab{\epsilon_{\alpha\beta}}
\def \Dab{\Delta_{\alpha\beta}}
\def \Dabn{\Delta_{\alpha\beta}^{\mbox{\tiny N}}}
\def \setDab{\left\{ \Delta_{\alpha\beta} \right\} }
\def \setDabn{\left\{ \Delta_{\alpha\beta}^{\mbox{\tiny N}} \right\} }
\def \li{\ell_i}
\def \leff{\ell_{\mbox{\tiny EFF}}}
\def \lj{\ell_j}
\def \si{s_i}
\def \sj{s_j}
\def \sm{s_m}
\def \be{\begin{equation}}
\def \ee{\end{equation}}
\def \bea{\begin{eqnarray}}
\def \eea{\end{eqnarray}}
\def \aQ{\alpha_{\mbox{\tiny Q}}}
\def \muQ{\mu_{\mbox{\tiny Q}}}
\def \lQ{\lambda_{\mbox{\tiny Q}}}
\def \lT{\lambda T}
\def \D{\partial}
\def \DF{\Delta F}
\def \DFdag{\Delta F^{\ddag}}
\def \dij{\delta_{ij}}
\def \Qdag{Q^{\dag}}
\def \Qddag{Q^{\ddag}}
\def \Qo{Q^{o}}
\def \En{E_{\mbox{\tiny N}}}
\def \dli{\delta\ell_i}
\def \dl{\delta\ell}
\def \dlj{\delta\ell_j}
\def \dei{\delta\epsilon_i}
\def \dsi{\delta s_i}
\def \de{\delta\epsilon}
\def \dej{\delta\epsilon_j}
\def \dQi{\delta Q_i}
\def \dQj{\delta Q_j}
\def \dQ{\delta Q}
\def \ds{\delta s}
\def \setei{\left\{ \epsilon_i \right\}}
\def \seteistar{\left\{ \epsilon_i^{\star} \right\}}
\def \setej{\left\{ \epsilon_j \right\}}
\def \setli{\left\{ \ell_i \right\}}
\def \setlj{\left\{ \ell_j \right\}}
\def \setQi{\left\{ Q_i \right\}}
\def \setQiQ{\left\{\Qi\left(Q\right)\right\}}
\def \Qistar{Q_i^{\star}}
\def \setQistarQ{\left\{ Q_i^{\star} \left(Q\right)\right\}}
\def \setQistar{\left\{ Q_i^{\star} \right\}}
\def \setQj{\left\{ Q_j \right\}}
\def \setQk{\left\{ Q_k \right\}}
\def \Ebar{\overline{E}}
\def \ebar{\overline{\epsilon}}
\def \eistar{\ei^{\star}}
\def \eibar{\overline{\epsilon_i}}
\def \libar{\overline{\ell_i}}
\def \ebarN{\overline{\epsilon}_{\mbox{\tiny N}}}
\def \en{\epsilon_{\mbox{\tiny N}}}
\def \lbar{\overline{\ell}}
\def \Tg{T_{\mbox{\tiny G}}}
\def \Tf{T_{\mbox{\tiny F}}}
\def \FH{F_{\mbox{\tiny HOMO}}}
\def \kH{k_{\mbox{\tiny HOMO}}}
\def \kB{k_{\mbox{\tiny B}}}
\def \wq{\omega_{\mbox{\tiny Q}}}
\def \Qast{Q^{\ast}}
\def \JQ{ {\cal J}^{\star}\left( Q\right)}

\renewcommand{\theequation}{\arabic{equation}}

\title{\Large\bf 
Investigation of routes and funnels in protein folding by free
energy functional methods
}
\author{\large Steven S.~Plotkin and Jos\'{e} N. Onuchic \\ 
Department of Physics, University of California, San Diego
}

\maketitle

%\newpage
{\small {\bf ABSTRACT}} $\;$ 
We use a free energy functional theory to elucidate general properties
of heterogeneously ordering, fast folding proteins, and we test our
conclusions with lattice simulations.
We find that both structural and energetic heterogeneity
can lower the free energy barrier to folding.
Correlating stronger contact energies with entropically likely contacts
of a given native structure lowers the barrier, 
and anticorrelating the energies has the reverse effect.
Designing in relatively mild energetic heterogeneity can eliminate the
barrier completely at the transition temperature.
Sequences with native energies tuned to  fold uniformly,
as well as sequences tuned to fold by a single or a few routes,
are rare. Sequences with weak native energetic heterogeneity are more common;
their  folding kinetics is more strongly determined by properties of
the native structure. Sequences with different distributions of
stability throughout the protein may still be good folders to the same
structure. A measure of folding route  
narrowness is introduced which correlates with rate, and which can give 
information about the intrinsic biases in ordering due to native
topology. This theoretical framework allows us to systematically investigate 
the coupled effects of energy and topology in protein folding, and to
interpret recent experiments which investigate these effects.

%\newpage
%{\small {\bf INTRODUCTION.}}$\;$
The energy landscape has been a central paradigm in understanding the
physical principles behind the self-organization of 
biological molecules~\cite{Onuchic97,DillKA97,VeitshansT97,GruebeleM99}. 
A central feature of landscapes of biomolecules which has emerged is
that the process of evolution, in selecting for sequences that fold
reliably to a stable conformation within a biologically relevant time,
induces a new energy 
scale into the
landscape~\cite{BryngelsonJD87,GoldsteinRA-AMH-92,ShakhnovichEI93a}.
In addition to the ruggedness energy scale already
present in heteropolymers, it now has the overall topography 
of a
funnel~\cite{LeopoldPE92,OnuchicJN95:pnas,PlotkinSS97,DillKA97}. A
sequence with a 
funneled landscape has a low energy native state  occupied with
large Boltzmann weight at temperatures high enough that folding
kinetics is not dominated by slow escape from individual traps. 

As an undesigned heteropolymer with a random, un-evolved sequence is
cooled, it becomes trapped into one of many structurally different low
energy states, similar to the phase transitions seen in 
spin glasses, glasses, and
rubber.
The low  temperature states typically look like a snapshot of the high
temperature collapsed states, but have dramatically slower
dynamics. 
On the other hand, when a designed heteropolymer or protein is cooled,
it reliably and quickly finds the dominant low energy structure(s)
corresponding to the native state, in a manner 
similar to the phase transition from the gas or liquid to the
crystal state. 
As in crystals,
the low temperature states typically have a lower symmetry
group than the many high temperature states~\cite{Wolynes96:symm}.
Connections have been made
between native structural symmetry and robustness to mutations of
proteins~\cite{Wolynes96:symm,LiH96,NelsonE97}. 
Funnel topographies are maximized in atomic clusters
when highly symmetric arrangements of the atoms are possible,
as in van der Waals clusters with ``magic
numbers''~\cite{WalesDJ99,BallKD96}, and similar 
arguments have been applied to proteins~\cite{Wolynes96:symm}, where
funneled landscapes are directly connected to mutational
robustness~\cite{PandeVS95:jcp}. 

It is appealing to make the connection between symmetry and
designability of native structures to the actual kinetics of the
folding process, arguing 
that symmetry or uniformity {\em in ordering} the protein maximizes the
number of folding routes and thus the ease of finding a candidate
folding nucleus, thus maximizing the folding rate.
Explicit signatures of multiple folding routes as
predicted by the funnel
theory~\cite{BryngelsonJD89,BryngelsonJD95}
have been seen in simulations of well-designed
proteins~\cite{LeopoldPE92,SaliA94:nat,BoczkoEM95,SocciND96:jcp,Lazaridis97,Pande99:pnas}
as well as experiments on several small globular
proteins~\cite{Oas97:nsb,Oliveberg98,Goldbeck99}. 
However these folding routes are not necessarily equivalent.
There is an accumulating body of 
experimental~\cite{FershtAR92,Radford92,BaiY95,SerranoL98nsb,BakerD98nsb}
and
simulation~\cite{AbkevichVI94,GutinAM95:pnas,PanchenkoAR96,Onuchic96,ShoemakerBA97,Lazaridis97,PortmanJprl98,Klimov98,Brookspnas98,Micheletti99,SheaJE99,NymeyerH00:pnas}
evidence which show varying degrees of heterogeneity in the
ordering process. These data refine the funnel picture by focusing on
which parts of the protein most effectively contribute to ordering,
and on the effects of native topology and native energy distribution
on rates and stability. 
The ensemble of foldable sequences with a given ratio of $\Tf/\Tg > 1$
has a wide distribution of mean first passage
times~\cite{BryngelsonJD89,GutinAM95:pnas,SocciND95:jcp},
indicating that several other properties of the
sequence and structure contribute to folding thermodynamics
and kinetics. These include  
topological properties of the native
structure~\cite{Wolynes96:symm,AbkevichVI95,BetancourtMR95,VigueraAR96,Plaxco98,ShoemakerWang99,AlmE99,MunozV99}
(e.g. mean 
loop length $\lbar$, dispersion in loop 
length $\delta \ell$, and kinetic accessibility of the native
structure), the distribution over contacts of  
total native energy in the protein, and the coupling of contact
energetics with native 
topology. 

In this paper we integrate the above sundry observations
into a theory which explicitly accounts for native heterogeneity,
structural and energetic, in the funnel picture.
We introduce a simple field theory with a
non-uniform order parameter to study fluctuations away from uniform
ordering, through free energy
functional methods introduced earlier by Wolynes and
collaborators~\cite{ShoemakerBA97,ShoemakerWang99}.~\footnote{
We treat only native couplings in detail, accounting for non-native
interactions as a uniform background field. 
Additionally, the correlation between contacts $(i,j)$ is a function only of
the overall order $Q$ in our theory.
This is analogous to the Hartree
approximation in the one-electron 
theory of solids~\cite{AndersonPW92}
where electrons mutually interact
only through an averaged field; extensions of our theory to include 
correlation mediated by native structure may be examined  within the
density-functional framework, and are a topic of future research. On
the other hand, tests of the theory by simulation %(fig.~(\ref{fig:conc})) 
(fig.~(1)) produce 
qualitatively the same results, so the conclusions are not
effected by including correlations to any order.}
The theory is in agreement with simulations also performed in this paper.
We organize the paper as follows. First we outline the calculation and
results. Next  we  derive and use an
approximate free energy functional 
which captures the essence of the problem. 
Then we conclude and suggest future research, leaving technical aspects
of the derivation for the methods section.

{\small {\bf OUTLINE.}}$\;$
The free energy functional description in principle allows for a
fairly complete understanding of the folding process for a particular
sequence; this includes effects due to 
the three dimensional topological native
structure, possible misfolded traps, and heterogeneity among the
energies of native contacts. We model a well-designed, minimally
frustrated  protein with an
approximate functional, but many of the results we obtain are quite
general. 
We find that for a well-designed protein, gains in loop
entropy and/or core energy  
always dominate over losses in route entropy, so the thermodynamic
folding barrier is 
always reduced by any preferential ordering in the
protein.~\footnote{
Folding heterogeneity effects the free energy in three ways: 1.) The
number of 
folding routes to the native state decreases; this effect increases
the folding barrier, 2.) The conformational entropy of polymer loops
increases, since native cores with larger halo entropies are more
strongly weighted. This decreases the folding barrier 3.) 
Making likely contacts stronger in energy lowers the 
thermal energy of partially native structures; this decreases the folding
barrier.}
However as long as ordering heterogeneity is not too large, there
are still many folding routes to the native structure, and the funnel
picture is valid. 
When there are very few routes to the native state due to large
preferential ordering, folding is slow and multi-exponential at
temperatures where the native structure is stable.
In this scenario the rate is governed by the kinetic
traps along the path induced, rather than the putative 
thermodynamic barrier which is absent.
Several physically motivated arguments giving the above results are
described in the supplementary material. 

To analyze the effects of native energetic as well as structural
heterogeneity on folding,
we coarsely describe the native structure through its distributions of
native contact energies $\setei$ and native loop lengths $\setli$. 
Here $\ei$ is the solvent averaged effective energy of contact $i$, and
$\li$ is the sequence length pinched off by contact $i$.
The labeling index $i$ runs from $1$ to $M$, where 
$M=z N$ is the total number of contacts, $N$ is the length of the
polymer, $z$ the number of contacts per residue. 
In the spirit of density functional theory of
fluids~\cite{PercusJK82} we introduce a coarse-grained 
free energy functional $F( \{
\Qi(Q)\} | \setei ,\setli )$ approximating the
physics of secondary (as e.g. along a helix) and tertiary
(non-local)  contacts in ordering. 
$Q$ is defined as the overall fraction of native contacts made,
used here to {\it stratify} the configurations with given similarity
to the 
native state, since this partitioning results in a funnel topography
of the energy landscape for designed
sequences~\cite{OnuchicJN95:pnas,PlotkinSS97}. 
The fraction of time contact $i$ is made in the
sub-ensemble of states at $Q$ is $\Qi(Q)$.
From a knowledge of this functional all relevant thermodynamic
functions can in general be calculated such as transition state
entropies and 
energies, barrier heights, and surface tensions. 
Moreover, derivatives of the functional give the
equilibrium distribution and correlation functions describing the
microscopic structure of the inhomogeneous system, as we see
below.

Given all the contact energies $\setei$ and loop lengths $\setli$ for
a protein, the thermal distribution of contact probabilities $\setQiQ$
is found by minimizing the free energy functional $F(\setQiQ | \setei,
\setli )$ subject to the constraint that the average probability is $Q$,
i.e. $\sum_i \Qi = M Q$ ($Q$ parameterizes the values of the $\Qi's$)~\footnote{
This procedure is analogous to finding the most probable distribution of
occupation numbers, and thus the thermodynamics, by maximizing the
microcanonical entropy for a system of particles obeying a given
occupation statistics - here the 
effective particles (the contacts) obey Fermi-Dirac statistics,
c.f. eq.~(\ref{eq:fd}).} 
Since in the model 
the probability of a contact to be formed is a function of its
energy and loop length, we can next consider the 
minimized free energy as  a function 
of the contact energies for a {\it given} native topology: $F(\setei |
\setli)$.  Then we can seek the special distribution of contact
energies $\{ \ei^\star (\li ) \}$ that minimizes or maximizes the
thermodynamic folding barrier to a particular structure by
finding the extremum of $F^\dag(\setei | \setli)$ with
respect to the contact energies $\ei$, subject to the
constraint of fixed native energy, $\sum_i \ei = M\ebar=\En $.
This distribution when substituted into the free energy 
gives in principle the extremum free energy barrier as a function of native
structure $F^\dag (\setli )$, which might then be optimized for the
fastest/slowest folding structure and its corresponding barrier.
We found that in fact the only distribution of energies for which the
free energy was an extremum is in fact the distribution which {\it
maximizes} the barrier by tuning all the contact probabilities to the
same value.

{\small {\bf METHODS.}}$\;$ 
We derive an approximate free energy functional, which takes account
for ordering heterogeneity, starting from a contact Hamiltonian 
${\cal H}( \{ \Dab \} | \{ \Dabn \}
)$ of the form
\be
{\cal H} = \sum_{\alpha<\beta} \left[ \eabn \Dab \Dabn + \eab\Dab
\left(1-\Dabn\right) \right]
\ee
Here the double sum is over residue indices, $\Dab=1$ ($0$) if residues
$\alpha$ and $\beta$ (do not) contact each other, $\Dabn=1$ ($0$) if
these residues (do not) contact each other in the native
configuration. The sum over native 
energies $\eabn$ and non-native energies $\eab$ gives the energy for a
particular configuration.~\footnote{A similar derivation of the free energy for
a uniform order parameter $Q$ was calculated in ref.~\cite{PlotkinSS97}.}
To obtain the thermodynamics we proceed 
by obtaining the distribution of state energies in
the microcanonical ensemble by averaging non-native interactions over
a Gaussian distribution of variance $b^2$:
$ P(E|\En, \{ \Dab\Dabn \} ) = 
\left<\right. \delta [ E- {\cal H} 
\{ \Delta_{\alpha\beta} \} ] \delta[ \En -
{\cal H} \{ \Delta_{\alpha\beta}^{\mbox{\tiny N}} \}
] \left. \right>_{n-nat} $.~\footnote{
This approach assumes minimal frustration, in that
native heterogeneity is explicitly retained and non-native
heterogeneity is averaged over;
phenomena specific to a
particular set of non-native energies, e.g. ``off-pathway''
intermediates, are smoothed over in this procedure.}
The averaging results in a Gaussian distribution having mean $\sum_i
\ei {\cal Q}_i$ and variance $Mb^2 (1-Q)$, where ${\cal Q}_i \equiv
\Dab\Dabn$ counts native contacts present in the configuration state
inside the stratum $Q$. From this distribution the log density of
states is obtained in terms of the configurational entropy of stratum
$Q$, $S(\{ {\cal Q}_i\} | Q )$, and the free energy functional 
$F(\{ {\cal Q}_i \} | Q)$ obtained
by performing the usual Legendre transform to the canonical
ensemble (c.f. eq~(\ref{F1})).~\footnote{Note that in eq.~(\ref{F1}) we
explicitly include the thermal trace over configurations at overall
order $Q$.}

We express
the free energy in terms of an arbitrary distribution of contact
probabilities - the distribution of $\{ \Qi \}$ that minimizes $F(\{ \Qi
\} | Q)$ is the (most probable) thermal 
distribution.~\footnote{In the contact representation, the averaged
bond occupation probabilities 
$\Qi=\left<\right. {\cal Q}_i \left. \right>_{\mbox{\tiny TH}}$ are
analogous to the averaged number density operator in an inhomogeneous
fluid:  
$\left< n({\bf x} ) \right>_{\mbox{\tiny TH}} = \left<\right. \sum_i \delta
({\bf x}_i -{\bf x} )  \left. \right>_{\mbox{\tiny TH}}$.}
For the ensemble of configurations at $Q$, we define the entropy that
corresponds to the multiplicity of contact patterns as 
${\cal S}_{\mbox{\tiny ROUTE}} ( \{ \Qi \} | Q)$
($>0$), and the configurational entropy lost from the coil state to
induce a contact pattern $\{ \Qi \}$ as
${\cal S}_{\mbox{\tiny BOND}} ( \{ \Qi \} | \setli , Q )$ ($<0$).
We make no capillarity or spinodal assumption, and treat the route entropy
as the entropy of 
a binary fluid mixture~\cite{PlotkinSS96,PlotkinSS97},
modified by a prefactor $\lambda(Q) \equiv
1-Q^{\alpha}$, which measures the number of 
combinatoric states reduced by chain topology: residues connected
by a chain have less mixing entropy than if they were 
free~\footnote{The value 
$\alpha=1.37$ gives the best fit to the lattice $27$-mer
data for the route entropy, while $\alpha\cong 1.0$ best fits the $27$-mer
free energy function. We generally use $\alpha\cong
1.0$ since the $27$-mer is small - for larger
systems $\alpha$ is smaller: more polymer is buried and thus more
strongly constrained by surrounding contacts.}:
\be
{\cal S}_{\mbox{\tiny ROUTE}} = 
\lambda\left(Q\right) \sum_{i=1}^{M} \left[ -Q_i \ln
Q_i - (1-Q_i)\ln\left(1-Q_i\right) \right] \: .
\label{smix}
\ee
We introduce a measure of ``routing'' ${\cal R}(Q)$ by expanding the
entropy to lowest order~\footnote{
We avoid the word ``pathway'' since  several definitions exist in the
literature; here a single route is unambiguously defined through the 
limit ${\cal S}_{\mbox{\tiny ROUTE}} \rightarrow 0$.}:
${\cal S}_{\mbox{\tiny ROUTE}} ( \{ Q+\dQi \}) \cong
{\cal S}_{\mbox{\tiny ROUTE}}^{\mbox{\tiny MAX}} - \lambda {\cal R}(Q)/2$,
where we have defined ${\cal R}(Q)$ by 
$
{\cal R}(Q) = \left< \dQ^2 \right> / \left< \dQ^2 \right>_{\mbox{\tiny
MAX}} \: ,
$
which is the variance of contact probabilities normalized by the
maximal variance,~\footnote{That is, if
$MQ$ contacts were made with probability $1$ and $M-MQ$ contacts were
made with probability $0$, then $\left<\right. (\Qi-Q)^2
\left. \right>_{\mbox{\tiny MAX}} = (1/M) (MQ (1-Q)^2 + (M-MQ) Q^2) =
Q(1-Q)$. Thus ${\cal R}(Q)$ is between $0$ and $1$.} 
In the limit 
${\cal R}(Q)=0$ the uniformly ordering system has the maximal route
entropy. When $\Qi=0$ or $1$ only, 
${\cal R}(Q)=1$, ${\cal S}_{\mbox{\tiny
ROUTE}}=0$, and only one route to the native state is 
allowed.~\footnote{That is, since all $\Qi$ are only zero or one at any
degree of nativeness,  
each successive bond added must always be the same one, so folding is then
a random-walk on the potential defined by that {\it single} route (there is
still chain entropy present). ${\cal R}(Q)$ is in the spirit of a
Debye-Waller factor applied to folding routes.}

In the supplementary material we derive a form for the configurational
entropy loss to fold to a given topological structure by accounting
for the distribution of entropy losses to form bonds or contacts due
to the distribution of sequence lengths in that structure.
We let the effective sequence (loop) length between residues $i$ and
$j$, $\leff (|i-j|, Q  )$ be a function of $Q$ (this is a mean field
approximation), and we take the entropy loss to close this loop to be
of the Flory form $\sim (3/2) \ln (a/\leff)$. The requirement that the
entropy be a state function restricts the possible functional form of
the effective loop length.
The result of the derivation for the contact entropy loss to form
state $\{\Qi\}$ is
\be
{\cal S}_{\mbox{\tiny BOND}} 
=  - (3/2) M \left( \left< \dQ\, \delta \ln
\ell \right> + S_{\mbox{\tiny MF}}(Q,\lbar) \right)
\label{Sfinal} 
\ee
where 
$\left<  \dQ \, \delta \ln \ell  \right> =
(1/M) \sum_i (\Qi -Q) (\ln \li - \overline{\ln \ell})$ is the
correlation between the  
fluctuations in contact probability and log loop length, 
and $S_{\mbox{\tiny MF}}(Q,\lbar)$ is the  mean-field bond
entropy loss (described in the supplement), and is a
function only of $Q$ and the mean
loop length $\lbar$.
By eq.~(\ref{Sfinal}) the
entropy is {\em raised} above that of a symmetrically ordering system
when shorter ranged contacts have higher probability to be formed;
this effect lowers the barrier.
Eq.s~(\ref{F1}), (\ref{smix}), and (\ref{Sfinal}) together give 
expression (\ref{Ffin}) for the free energy 
$F( \{ \Qi(Q)\} | \setei ,\setli )$ of a well-designed protein
that orders heterogeneously.

The lattice protein used in fig~1 to check the theory is a
chain of $27$ monomers constrained to the vertices of a 3-D cubic
lattice. Details of the model and its behavior can be found
in~\cite{LeopoldPE92,SaliA94:nat,AbkevichVI94,SocciND95:jcp,SocciND96:jcp,NymeyerH00:pnas}.
Monomers have non-bonded contact interactions with a G\={o} potential
(native interactions only).~\footnote{Corner, crankshaft, and end moves are
allowed. Free energies and contact probabilities are obtained by
equilibrium monte-carlo sampling using the histogram
method~\cite{SocciND95:jcp}. Sampling error is $< 5 \%$.}
Coupling energies were chosen for row 1 of fig 1 by first running a
simulated annealing algorithm to find the set $\{ \ei^\star \}$ that makes
all the $\Qi(\{ \ei^\star \}) = \Qddag$ at the barrier peak. Energies
are always constrained to sum to a fixed total native energy: $\sum_i
\ei = M \ebar$. Then
energies were relaxed by letting  
$\ei = \eistar + \alpha ( \eibar - \eistar )$. The values $\alpha = 1$, $
1.35$, $2.05$ were used in rows 2, 3, and 4 respectively.

{\small {\bf FREE ENERGY FUNCTIONAL.}}$\;$ 
By averaging a contact Hamiltonian over non-native interactions, we
can derive  an approximate free energy
functional for a well-designed protein (See the methods section). 
We analyze here heterogeneity in minimally frustrated
sequences, where the roughness energy scale $b$ is smaller than 
the stability gap $\ebar$.
The general form of the free energy functional is
\be
F = \!\left<
\sum_{i=1}^{M} \left[ \ei {\cal Q}_i - T S\left(\{ {\cal
Q}_i\} | Q \right) \right]  \right>_{\!\mbox{\tiny THERM}}^{'}
\!\!\!\!\!\!\!- \frac{M b^2}{2T} \left(1-Q\right) 
\label{F1}
\ee
where ${\cal Q}_i=(0,1)$ counts native contacts in a configurational
state (so the sum on $\ei {\cal Q}_i$ gives the states energy),
summing
$S(\{ {\cal Q}_i\} | Q )$ gives the states configurational entropy,
and then this is thermally averaged over all states restricted to have
$MQ$ contacts. The second term accounts for low energy non-native traps. 

The study of the configurational entropy is a fascinating but
complicated problem detailed in the methods section.
In summary this entropy functional generalizes the Flory mean-field
result~\cite{FloryPJ56:jacs,PlotkinSS96} to account for 
the topological heterogeneity inherent in the native structure and
a finite average return length for that structure 
(contact order~\cite{Plaxco98}), 
as well as to account for the number of folding routes to the native
structure. The amount of route diversity or narrowness in folding can be
quantified in terms of the 
relative fluctuations of contact formation $\dQ = \Qi(Q)-Q$:
\be
{\cal R}(Q) = \left< \dQ^2 \right> / \left< \dQ^2 \right>_{\mbox{\tiny
MAX}} \: ,
\label{eq:routem}
\ee
which is useful for our analysis below.
Our resulting analytic expression for the free energy of a protein 
that folds heterogeneously is~\footnote{We have expanded the route
entropy eq.~(\ref{smix}) to 
		  second order in this expression for clarity; in
		  deriving the results of the theory the full
		  expression is used.}.
\be
\frac{F}{M} \!\cong\! \frac{F_{\mbox{\tiny MF}}^o}{M} \! + \!
\overline{\dQ \, \de}  + \frac{\lT}{2} \frac{\overline{\dQ^2}}{Q
\left(1\! -\! Q\right)}  + \frac{3}{2} \overline{ \dQ \, \delta \ln \ell
} \, .
\label{Ffin}
\ee
Here $F_{\mbox{\tiny MF}}^o (Q)$ is the uniform-field free energy
function (similar to that obtained previously
in~\cite{PlotkinSS97}). The free energy functional is approximate
in that it results from an integration over a local free 
energy density whose only information about 
the surrounding medium is through the average field present ($Q$), 
$F = \sum_i f_i(\Qi,Q)$. Cooperative entropic effects due to local
correlations~\cite{DillKA93:pnas,ShoemakerWang99}
between 
contacts would be an important extension of the model, and have been
treated elsewhere in similar models~\cite{ShoemakerWang99}.
Inspection of eq.~(\ref{Ffin}) shows that as heterogeneity increases,
the effect on the barrier is a competition between energetic and
polymer entropy gains (2nd and 4th terms) and route entropy losses
(3rd term) as described above.

Minimizing the free energy~(\ref{Ffin}) at fixed $Q$, $\delta( F +
\mu\sum_j \Qj ) = 0$, gives a Fermi distribution for the most probable
bond occupation probabilities $\setQistar$ for a given $\setei$ and
$\setli$: 
\be
\Qi^{\star}(Q) = 1/\left( 1 + \exp\left[  \left(\mu' + \ei
-T \si \right)/\lT \right] \right)
\label{eq:fd}
\ee
where the Lagrange multiplier $\mu' \sim -(1/M) \D F/\D Q$ is related
to the effective force on the potential $F(Q)$. Positive second
variation of $F$ indicates the extremum is in fact a minimum.

{\small {\bf OPTIMIZING RATES, STABILITY, AND ENTROPY}}$\;$ 
We now consider the effects on the free energy when the native 
interactions between residues are changed in a controlled manner.
The theory
predicts a barrier at the transition temperature 
of a few $\kB T$, in general agreement with
experiments on small, single-domain proteins. The barrier height is fairly  
small compared to the total thermal energy of the system, reflecting the
exchange of entropy for energy as the protein folds.
However the barrier height can vary significantly depending on which
parts of the protein are more stable in their local native structure.
At uniform stability we find the largest barrier (for a given total
native energy): about twice as large
as the barrier when stability is governed purely by the three-dimensional
native structure, i.e. when all interaction energies are equal.
Increasing heterogeneity, by energetically favoring regions of the
protein which are already entropically likely to order, systematically
decreases the barrier, and in
fact can eliminate the barrier entirely if the heterogeneity is large
enough. See figure~1. %figure~\ref{fig:conc}.

We seek to relax the values of $\setej$ at fixed native energy $\En =
\sum_j \ej$ to the distribution $\left\{
\ei^{\star}\left(\setlj\right)\right\}$ that extremizes the free energy
barrier, by finding the solution of $\sum_i [ \delta
F^{\ddag}/\delta\ei -p]\dei =0 $ for arbitrary and independent
variations $\dei$ in the energies. It can be shown
that $\delta F /\delta\ei = \D F/\D \ei + \mu (\delta/\delta\ei)\sum_j
\Qj$, however the second term is zero since $\delta Q/\delta\ei =0$,
so by eq.~(\ref{F1}) $\delta F /\delta\ei = \Qi$: 
the contact probability plays the role of the local density, and the
perturbation $\dei$ the role of an external field, as in liquid state
theory. 
At the extremum
all contact probabilities are equal: $\Qi = p = \Qdag$, which in our
model means that longer loops have lower (stronger) energies: $\dei =
T \dsi = -(3/2) T \, \delta \ln \li $; there is full symmetry in the
ordering of the protein at the extremum. 
Evaluating the second derivative mechanical-stability matrix shows 
$\Qi = \Qddag$  to be an unstable maximum:
\be
\left(\delta^2 F^{\ddag}/\delta\ej\delta\ei\right)_{\ei^{\star},\ej^{\star}} = -
\dij \;
\Qddag \left(1-\Qddag\right)/\lambda^{\ddag} T  \: .
\label{2nd}
\ee
This is clearly negative, meaning that tuning the energies so that
$\Qi = \Qddag$ {\it maximizes} the free energy at the barrier
peak. Since the change in the unfolded state (at $Q \approx 0$) is much
weaker than at the transition state, 
the barrier height itself is essentially maximized.
Substituting eq.~(\ref{2nd}) into a Taylor expansion of the free
energy at the extremum (and using $\lambda^{\ddag}=\lambda(\Qddag)
\approx 1-\Qddag$) gives for the rate 
\be
k = k_{\mbox{\tiny HOMO}} 
\exp ( \Qddag M \overline{\de^2}/2 T^2 ) \: ,
\label{eq:k}
\ee
which is to be compared
with eq.~(1) in the supplementary material (obtained by an argument
using the random energy model).
In terms of the route narrowness 
measure ${\cal R}(Q)$ the change in free
energy barrier on perturbation is
\be
\delta \DFdag = - (1/2) M \lambda^{\ddag} T \, {\cal R}(\Qddag) \: .
\label{dfhet}
\ee
A variance in contact participations $\overline{\dQ^2} = 0.05$ which
is about $20\%$ of the maximal dispersion 
($\approx 1/4$, taking $\Qddag\approx 1/2$) lowers the
barrier by about $0.1 N\kB T$ or about $5 \kB T$ for a chain of
length $N \approx 50$ (believed to model a 
protein with $\sim 100$~aa~\cite{OnuchicJN95:pnas}).  

We can extend the analysis by perturbing about a
structure with mean loop length $\lbar$, and including effects on the
barrier due to dispersion in loop length and correlations between
energies and loop lengths. A perturbation expansion of the free energy
gives to lowest order:
\be
\frac{\delta \Delta F^{\ddag}}{M} = 
- \frac{\Qddag}{2 T}  \overline{\de^2}
- T \frac{9}{8} \Qddag
\frac{\overline{\dl^2} }{\lbar^2}
- \frac{3}{4} \Qddag \frac{\overline{\dl \de}}{\lbar}
\label{pertF}
\ee
indicating that the free energy barrier is additionally lowered 
by structural {\it variance} in loop lengths, and also
when shorter range 
contacts become stronger energetically ($\dli <0$ and $\dei <0$) or
longer range contacts become weaker energetically ($\dli >0$ and
$\dei >0$) i.e. in the model the free energy is additionally lowered
when fluctuations are correlated so as to further increase the variance in
contact participations. This effect has been seen in
experiments by the Serrano group~\cite{VigueraAR96,Munoz96:Rev}. 

To test the validity of the theory, we compare the analytical
results obtained from our theory with the results from simulation of a
27-mer lattice protein model. The comparison is shown on figure 1
where a full analysis is performed. All energies are in units of the mean
native interaction strength $\ebar$.

The rate dependence on heterogeneity 
should be experimentally testable by measuring the
dependencies of folding rate at the transition temperature of a
well-designed protein on the 
{\it dispersion} of $\phi$-values. It is important that before and after the
mutation(s) the protein remains fast-folding to the native structure 
without ``off-pathway'' intermediates, and that its native state
enthalpy remain approximately the same, perhaps by tuning
environmental variables.

{\small {\bf CONCLUSIONS AND FUTURE WORK.}}$\;$ 
In this paper we have introduced refinement and insight into the
funnel picture by considering heterogeneity in the folding of
well-designed proteins. 
We have explored in minimally frustrated sequences how folding is
effected by 
heterogeneity in native contact energies, as well as the entropic
heterogeneity inherent in folding to a specific 
three-dimensional native structure.
Specifically we examined the effects on the folding free energy barrier,
distribution of participations in the transition state ensemble
TSE',~\footnote{We use a prime since we actually look at the barrier
peak along the $Q$ coordinate.} as well 
as the diversity or narrowness of folding routes.
For the ensemble of sequences having a given $\Tf/\Tg$,
homogeneously ordering sequences have the largest folding
free energy barrier.
For most structures, where
topological factors play an important role, this regime is achieved by
introducing a large dispersion in the distribution of native contact
energies which in practice would be almost impossible to achieve. As we
reduce the dispersion in the contact energy distribution to a uniform
value $\ebar$, the dispersion of contact participations increases and
thus the number of folding routes decreases, the free
energy barrier decreases and the total configurational entropy at
the TSE' initially increases due to polymer halo effects. 
The folding temperature is only mildly effected; the prefactor
appearing in the rate is probably only
mildly effected also, since it is largely a function of $\Tf/\Tg$ and
polymer properties~\cite{SocciND96:jcp}.
Tuning the interaction energies further results in more probable
contacts having stronger energy. Route diversity decreases to moderate
values - there are still many routes to the native state, and
$\Tf/\Tg$ is still sufficiently greater than one. The 
barrier eventually decreases to zero, at relatively mild dispersion in
native contact energy.
The funnel picture, with different structural details, is valid for
the above wide range of native contact energy distributions. 
However, tuning the energies further so that probable contacts have even lower
energy eventually induces the system to take a single or very few
folding routes at the transition temperature.
A large dispersion of energies is required to achieve this,
and in this regime the folding temperature drops well below
the glass temperature range, where folding rates are extremely slow.

Since fine tuning interactions on the funnel may
effect the rate, sequences may be designed to fold both faster or slower
to the {\it same} structure of a wild type sequence, depending how the
interaction strengths correlate with the entropic likelihood of contact
formation.  Folding rates in mutant proteins that exceed those of the
wild type have been receiving much interest in recent 
experiments~\cite{VigueraAR96,Munoz96:Rev,HagenSJ96:pnas,KimDE98,BrownBM99}.
Enhancement (or suppression) of folding rate to a given structure due
to changes in sequence are modeled in our theory through changes in
native interactions; our results are fully supported by the
experiments cited above. The fact that a minimally frustrated protein
is robust to perturbations in the interactions
means that at least the folding scenarios depicted in the center 2
rows of fig.~1 are feasible within the ensemble of sequences
that fold to the given structure. However the number of sequences should
be maximal when all the native interactions are near their
average, and the actual width of the native interactions depends on the
true potential energy function. Fluctuations in rate due to the weakening
or strengthening of non-native traps by sequence perturbations is an
interesting topic of future research.
The enhancements or reductions in rate we have explored are 
mild compared to the enhancement by minimal frustration (funneling the
landscape):
the fine tuning of rates may be a phenomenon manifested by {\it in
vitro} or {\it in machina} evolution, rather than {\it in vivo}
evolution. Nevertheless rate tuning and folding heterogeneity 
may become an important factor for larger
proteins, 
where e.g. stabilizing partially native intermediates may increase the
overall rate or prevent aggregation.
Given that a sequence is minimally frustrated, 
heterogeneity or broken-ordering-symmetry in
fact aids folding, similar to the enhancement of nucleation rates
seen in other disordered media~\cite{Oxtoby96}. 
Similar effects have been observed in Monte Carlo simulations
of sequence evolution, when the selection criteria involves fast
folding rate~\cite{GutinAM95:pnas}. Here
we see how such phenomena can arise from general considerations of the
energy landscape theory.
The notion that rates increase with heterogeneity at little expense to
native stability contrasts with the 
view that non-uniform ordering exists
merely as a residual signature of incomplete evolution to a
uniformly folding state. 
Adjusting the backbone rigidity or the non-additivity of
interactions~\cite{PlotkinSS97,KolinskiA93:jcp} can also modify the
barrier height, 
possibly as much as the effects we are considering here.
There may also be functional reasons for non-uniform folding - 
malleability or rigidity requirements of the active
site may inhibit or enhance its tendency to order.
The amount of route narrowness in folding was introduced as a
thermodynamic measure through
the mean square fluctuations in a local order parameter. 
The route measure may be useful in
quantifying the natural kinetic accessibility of various structures.
While structural heterogeneity is essentially always present,
the flexibility inherent in the number of
letters of the sequence code limits
the amount of native energetic heterogeneity possible.
%~\cite{PlotkinThesis},
However some sequence flexibility is in fact required for funnel
topographies~\cite{Wolynes97nsb} and so is probably present
at least to a limited degree. 
We have seen here how a very general theoretical framework can be
introduced to explain and understand the effects of local
heterogeneity in native stability and structural topology on such
quantities as folding rates, transition temperatures, and the degree
of routing in the funnel folding mechanism. Such a theory should be a
useful guide in interpreting and predicting experimental results on
many fast-folding proteins.

We thank Peter Wolynes, Hugh Nymeyer, Cecilia Clementi, and Chinlin Guo
for their generous and insightful discussions. This work was
initiated while Plotkin was a graduate student with Peter Wolynes.
This work was supported by NSF Grant MCB9603839 and NSF Bio-Informatics
fellowship DBI9974199.

\newpage
\onecolumn
\vspace{-1.0cm}
\begin{figure}[htb]
\hspace{.7cm}
\centerline{\psfig{file=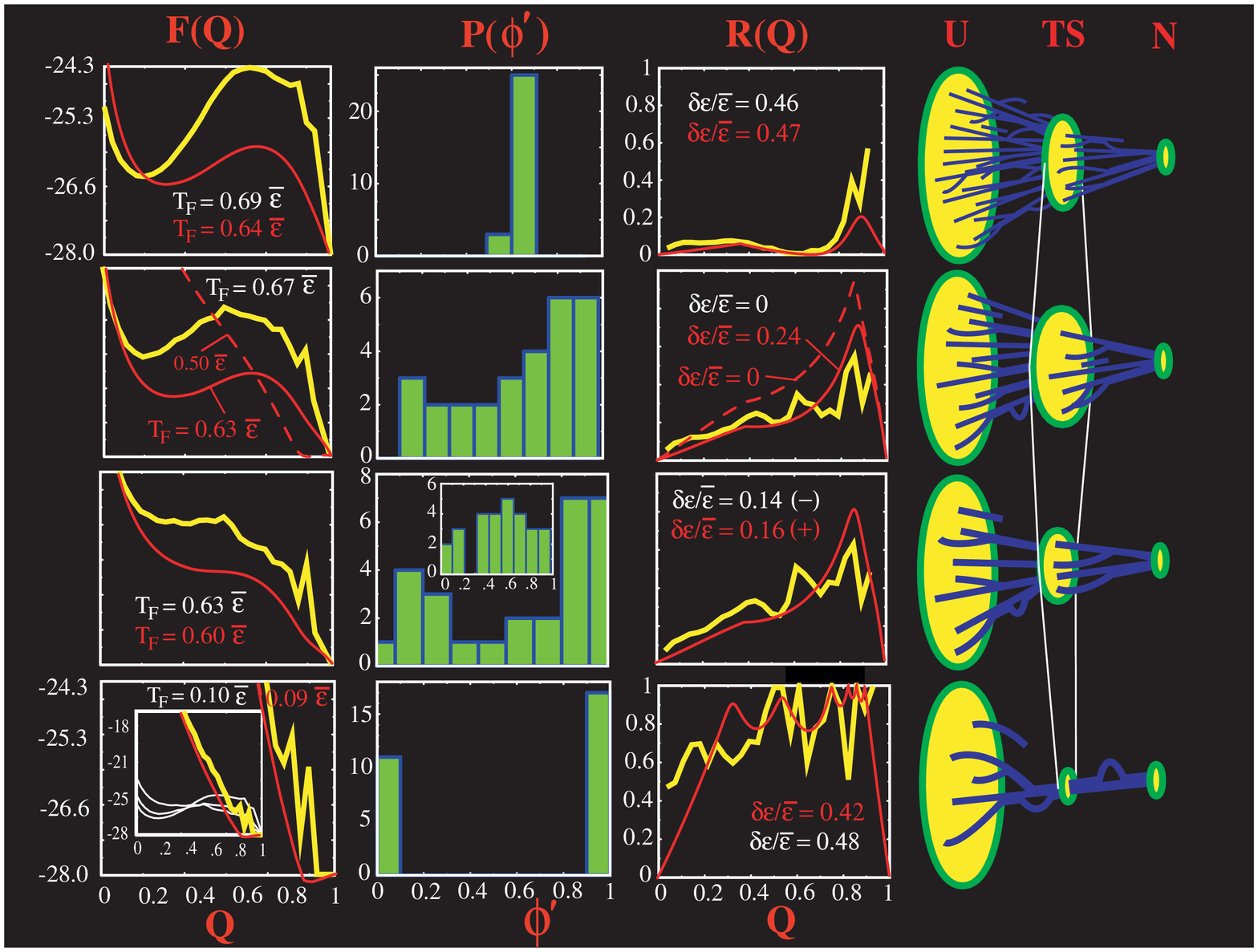,height=18cm,width=14cm,angle=-90}}
\caption{}
\label{fig:conc}
\end{figure}

\newpage
%\baselineskip 0.33cm
%{ \small 
{\bf CAPTION FOR FIG.~1:} \\
The effects of heterogeneity in contact probability (increased from top
to bottom) on barrier height $F^{\ddag}$, folding temperature $\Tf$, and
ordering heterogeneity are
summarized here; plots are for simulations of a $27$-mer lattice G\={o}
model (yellow) to the {\it same} native structure (given
in~\cite{SocciND96:jcp}), and for 
the analytic theory in the text (red).
The simulation results make no assumptions on the nature of the
configurational entropy; the theoretical results use the approximate
state function of eq.~(\ref{Sfinal}), along with a
cutoff used for the shorter loops so the bond entropy loss for each
loop is always $\leq 0$ (the same loop length distribution as in the
lattice structure is used).
In the top row, energies are tuned for both simulation and theory
to fully symmetrize the funnel: $\Qi(\ei^{\star})=Q$; Second row:
energies are then relaxed for the simulation results so they are all
equal: $\ei = \ebar$; energies in the theory are relaxed the same way
until a comparable $\Tf$ is achieved; Third 
row: energies are then further tuned to a distribution $\ei \cong
\ei^o$ that kills the barrier 
(there a many such distributions - all that is necessary is sufficient
contact heterogeneity); 
The top 3 rows are funneled folding mechanisms with many routes to
the native structure. 
Last row: energies are tuned to induce a
single or a few specific routes for folding.
All the while the energies are constrained to sum to $\En$: $\sum_i
\ei = \En$.
The free energy profile $F(Q)$ (in units of $\ebar$) 
is plotted in the left column at the
folding  transition temperature $\Tf$, which is given. 
The next column shows the distribution of
thermodynamic contact probabilities $\Qi(\Qddag) \equiv \phi'$ at the
barrier peak (we use the notation $\phi'$  since this is a
thermodynamic rather than kinetic measurement, however for
well-designed proteins the two are strongly correlated  with coefficient
$\approx 0.85$~\cite{NymeyerH00:pnas}). Only simulation 
results are shown to keep the figure easy to read; the theory gives $\phi'$
distributions within $\sim 10\%$ as may be inferred from their similar
route measures.
The next column shows the route measure ${\cal R}(Q)$ of
eq.~(\ref{eq:routem}) and gives the
dispersion in native energies required to induce the scenario of that
row (${\cal R}(0,1)=0/0$ is undefined and so is omitted from the
simulation plots; it is defined in the theory through the limit
$Q\rightarrow 0,1$).
The right column  shows schematically the different folding routes 
as heterogeneity is increased; from a maximum number of routes through
$\Qdag$ to essentially just one route.
{\bf TOP ROW}: 
In the uniformly ordering funnel we can see first that $P(\phi')$ is
a delta function and ${\cal
R}(\Qddag) = 0$ (c.f. eq.~(\ref{eq:routem})), so ordering at the
transition state (or barrier peak $\Qdag$) is essentially
homogeneous. The number of 
routes through the 
bottleneck (c.f. eq.~(\ref{smix})) is maximized, as schematically drawn on
the right. Branches are drawn in the routes to illustrate the
minimum of ${\cal R}(Q)$ at $\Qddag$.
The free energy barrier is maximized (eq.~(\ref{dfhet})), thus
the stability of the native state at fixed temperature and
native energy is maximized, and so the
folding temperature $\Tf$ at fixed native energy is maximized. 
$\Tf$ in the  simulation is defined as the temperature where the
native state ($Q=1$) is occupied $50\%$ of the time. In the theory,
at $\Tf$ the probability for $Q \geq 0.8$ is $0.5$.
A very large dispersion in energies is required to induce this
scenario; some contact energies are nearly zero, others are several
times stronger than the average.
{\bf SECOND ROW}: 
In the uniform native energy funnel the barrier height is roughly
halved while hardly changing $\Tf$, for the following reason.
In a G\={o} model, as the contact energies are relaxed from
$\{\eistar\}$ to a 
uniform value $\ei =\ebar$, the energy of the transition state is essentially
constant: initially the energy is $\sum_i \Qistar(\Qdag) \eistar = Q
\sum_i \eistar = Q\En$, and as the contact energies are relaxed to a
uniform value $\sum_i \Qi \ebar = \ebar \sum_i \Qi = Q\En$ once again.
However the transition state entropy increases and obtains its
maximal value when $\ei=\ebar$, because then all microstates at
$\Qdag$ are equally probable since the probability to occupy a
microstate is $p_i \sim
\exp(-E_i(\Qdag)/T)=\exp(-Q\En/T)/Z = 1/\Omega(\Qdag)$. 
The thermal entropy $-\sum_i p_i \log 
p_i$ then equals the configurational entropy $\log \Omega(\Qdag)$ (its
largest possible value). Thus as
contact energies are relaxed from $\eistar$
where they are anti-correlated to their loop
lengths (more negative energies tend to be required for longer loops 
to have equal free energies) 
to  $\ebar$ where they are uncorrelated to their
loop lengths, the barrier initially decreases because the total
entropy of the bottleneck increases (drawn schematically on the
right), i.e. increases in polymer halo entropy 
are more important than decreases in route entropy. The system is
still sufficiently two-state that $\Tf$ is hardly changed.
$P(\phi')$ is broad indicating inhomogeneity in 
the transition state, due solely in this scenario to the topology of
the native 
structure since all contacts are equivalent energetically; Routing is
more pronounced - when $\ei=\ebar$, ${\cal R}(Q)$  is measure of
the intrinsic fluctuations in order due to the natural inhomogeneity
present in the native structure; different structures will have
different profiles and it will be interesting to see how this measure
of structure couples with thermodynamics and kinetics of folding.
Loops and dead ends in the schematic drawings are used to illustrate
local decreases and increases in ${\cal R}(Q)$; these fluctuations are
captured by the theory only when the routing becomes pronounced (last row).
The solid curves presented for the theory are shown for a 
reduction in $\Tf$ comparable to the simulations. There is still
some energetic heterogeneity present as indicated. When $\ei =\ebar$
in the theory 
(dashed curves), the fluctuations in $\Qi$ are somewhat larger than the
simulation values, and the entropic heterogeneity is sufficient to
kill the barrier- the free energy is downhill at $\Tf \cong 0.5
\ebar$. The free energy  barrier results from a cancellation
of large terms and is significantly more sensitive than intensive
parameters such as route measure ${\cal R}(Q)$.
{\bf THIRD ROW}: 
In approaching the zero-barrier funnel scenario for the simulation,
the energies are 
further perturbed and now begin to 
anti-correlate with contact probability (and tend to correlate with
loop length); i.e. more probable contacts (which tend to have shorter loops)
have stronger energies. For the theory not as much heterogeneity is
required. Contact energies are still correlated with formation
probability as  indicated by the signs in parentheses.
The free energy barrier continues to decrease until some set
of energies $\{ \eio \}$ where the barrier at $\Tf$ vanishes entirely.
All the while the transition temperature $\Tf$ decreases only $\sim 10\%$,
so that slowing of dynamics (as $\Tf$ approaches $\Tg$) would not be a major
factor.
At this point the $\phi'$ distribution at the barrier position
$\Qddag(\ebar)$ is essentially bi-modal, but the distribution at
$\Qddag(\{\ei^o\})$ (inset) is less so because of transition state drift
towards lower $Q$ values (the Hammond effect). A relatively small amount of
energetic heterogeneity is needed to kill the barrier at $\Tf$. There
are still many routes to the native state since ${\cal R}(\Qddag)
\approx 0.3-0.4$, but some contacts are fully formed in the transition
state (some $\phi' \cong 1$).
{\bf BOTTOM ROW}: 
As the energies continue to be perturbed to values that cause
folding to occur by a single dominant route rather than a funnel 
mechanism, folding becomes strongly downhill
at the transition temperature, which drops
more sharply towards $\Tg$: here to induce a single pathway $\Tf$ must
be decreased to about $1/4$ 
the putative estimate of $\Tg$ (about $\Tf(\{ \ebar \})/1.6$,
see~\cite{OnuchicJN95:pnas}).
In this scenario, the actual shape of the free energy profile depends
strongly on which route the system is tuned to; Non-native
interactions not included here become important.
Contact participation at the barrier is essentially one or zero, and
the route measure at the barrier is essentially one. The entropy at
the bottleneck is relatively small (the halo entropy of a single
native core). The energetic heterogeneity necessary to achieve
this scenario is again very large - comparable to what is needed to
achieve a uniform funnel. 
%}

\twocolumn

\newpage
\pagestyle{plain}

%\nocite{TitlesOn}
%\bibliography{plotkin/TeX/Bib/savenL}
%\bibliography{savenbib}			%~/TeX/Bib/saven.bib}

\begin{thebibliography}{10}

\bibitem{Onuchic97}
Onuchic, J.~N, Luthey-Schulten, Z,  \& Wolynes, P.~G.
\newblock (1997) {\em Annu Rev Phys Chem} {\bf 48}, 545--600.

\bibitem{DillKA97}
Dill, K.~A \& Chan, H.~S.
\newblock (1997) {\em Nat. Struct. Biol.} {\bf 4}, 10--19.

\bibitem{VeitshansT97}
Veitshans, T, Klimov, D,  \& Thirumalai, D.
\newblock (1997) {\em Folding and Design} {\bf 2}, 1--22.

\bibitem{GruebeleM99}
Gruebele, M.
\newblock (1999) {\em Annu Rev Phys Chem} {\bf 50}, 485--516.

\bibitem{BryngelsonJD87}
Bryngelson, J.~D \& Wolynes, P.~G.
\newblock (1987) {\em Proc Nat Acad Sci USA} {\bf 84}, 7524--7528.

\bibitem{GoldsteinRA-AMH-92}
Goldstein, R.~A, Luthey-Schulten, Z.~A,  \& Wolynes, P.~G.
\newblock (1992) {\em Proc Nat Acad Sci USA} {\bf 89}, 4918--4922.

\bibitem{ShakhnovichEI93a}
Shakhnovich, E.~I \& Gutin, A.~M.
\newblock (1993) {\em Proc Nat Acad Sci USA} {\bf 90}, 7195--7199.

\bibitem{LeopoldPE92}
Leopold, P.~E, Montal, M,  \& Onuchic, J.~N.
\newblock (1992) {\em Proc Nat Acad Sci USA} {\bf 89}, 8721--8725.

\bibitem{OnuchicJN95:pnas}
Onuchic, J.~N, Wolynes, P.~G, Luthey-{S}chulten, Z,  \& Socci, N.~D.
\newblock (1995) {\em Proc Nat Acad Sci USA} {\bf 92}, 3626--3630.

\bibitem{PlotkinSS97}
Plotkin, S.~S, Wang, J,  \& Wolynes, P.~G.
\newblock (1997) {\em J Chem Phys} {\bf 106}, 2932--2948.

\bibitem{Wolynes96:symm}
Wolynes, P.~G.
\newblock (1996) {\em Proc Nat Acad Sci USA} {\bf 93}, 14249--14255.

\bibitem{LiH96}
Li, H, Helling, R, Tang, C,  \& Wingreen, N.
\newblock (1996) {\em Science} {\bf 273}, 666--669.

\bibitem{NelsonE97}
Nelson, E.~D, Teneyck, L.~F,  \& Onuchic, J.~N.
\newblock (1997) {\em Phys Rev Lett} {\bf 79}, 3534--3537.

\bibitem{WalesDJ99}
Wales, D.~J \& Scheraga, H.~A.
\newblock (1999) {\em Science} {\bf 285}, 1368--1372.

\bibitem{BallKD96}
Ball, K.~D, Berry, R.~S, Kunz, R.~E, Li, F.~Y, Proykova, A.~A,  \& Wales, D.~J.
\newblock (1996) {\em Science} {\bf 271}, 963--966.

\bibitem{PandeVS95:jcp}
Pande, V.~S, Grosberg, A.~Y,  \& Tanaka, T.
\newblock (1995) {\em J Chem Phys} {\bf 103}, 9482--9491.

\bibitem{BryngelsonJD89}
Bryngelson, J.~D \& Wolynes, P.~G.
\newblock (1989) {\em J Phys Chem} {\bf 93}, 6902--6915.

\bibitem{BryngelsonJD95}
Bryngelson, J.~D, Onuchic, J.~N, Socci, N.~D,  \& Wolynes, P.~G.
\newblock (1995) {\em Proteins} {\bf 21}, 167--195.

\bibitem{SaliA94:nat}
{\u{S}}ali, A, Shakhnovich, E,  \& Karplus, M.
\newblock (1994) {\em Nature} {\bf 369}, 248--251.

\bibitem{BoczkoEM95}
Boczko, E.~M \& Brooks, C.~L.
\newblock (1995) {\em Science} {\bf 269}, 393--396.

\bibitem{SocciND96:jcp}
Socci, N.~D, Onuchic, J.~N,  \& Wolynes, P.~G.
\newblock (1996) {\em J Chem Phys} {\bf 104}, 5860--5868.

\bibitem{Lazaridis97}
Lazaridis, T \& Karplus, M.
\newblock (1997) {\em Science} {\bf 278}, 1928--1931.

\bibitem{Pande99:pnas}
Pande, V.~S \& Rokhsar, D.~S.
\newblock (1999) {\em Proc Nat Acad Sci USA} {\bf 96}, 1273--1278.

\bibitem{Oas97:nsb}
Burton, R.~E, Huang, G.~S, Daugherty, M.~A, Calderone, T,  \& Oas, T.~G.
\newblock (1997) {\em Nature Struct Biol} {\bf 4}, 305--310.

\bibitem{Oliveberg98}
Oliveberg, M, Tan, Y, Silow, M,  \& Fersht, A.
\newblock (1998) {\em J Mol Biol} {\bf 277}, 933--943.

\bibitem{Goldbeck99}
Goldbeck, R.~A, Thomas, Y.~G, Chen, E, Exquerra, R.~M,  \& Kligar, D.~S.
\newblock (1999) {\em Proc Nat Acad Sci USA} {\bf 96}, 2782--2787.

\bibitem{FershtAR92}
Fersht, A.~R, Matouschek, A,  \& Serrano, L.
\newblock (1992) {\em J Mol Biol} {\bf 224}, 771--782.

\bibitem{Radford92}
Radford, S.~A, Dobson, M, \& Evans, P.~A.
\newblock (1992) {\em Nature} {\bf 358}, 302--307.

\bibitem{BaiY95}
Bai, Y, Sosnick, T.~R, Mayne, L,  \& Englander, S.~W.
\newblock (1995) {\em Science} {\bf 269}, 192--197.

\bibitem{SerranoL98nsb}
Martinez, J.~C, Pisabarro, M.~T,  \& Serrano, L.
\newblock (1998) {\em Nature Struct Biol} {\bf 5}, 721--729.

\bibitem{BakerD98nsb}
Grantcharova, V.~P, Santiago, J.~V, Baker, D,  \& Riddle, D.~S.
\newblock (1998) {\em Nature Struct Biol} {\bf 5}, 714--720.

\bibitem{AbkevichVI94}
Abkevich, V.~I, Gutin, A.~M,  \& Shakhnovich, E.~I.
\newblock (1994) {\em Biochemistry} {\bf 33}, 10026--10036.

\bibitem{GutinAM95:pnas}
Gutin, A.~M, Abkevich, V.~I,  \& Shakhnovich, E.~I.
\newblock (1995) {\em Proc Nat Acad Sci USA} {\bf 92}, 1282--1286.

\bibitem{PanchenkoAR96}
Panchenko, A.~R, Luthey-Schulten, Z,  \& Wolynes, P.~G.
\newblock (1996) {\em Proc Nat Acad Sci USA} {\bf 93}, 2008--2013.

\bibitem{Onuchic96}
Onuchic, J.~N, Socci, N.~D, Luthey-Schulten, Z,  \& Wolynes, P.~G.
\newblock (1996) {\em Folding and Design} {\bf 1}, 441--450.

\bibitem{ShoemakerBA97}
Shoemaker, B.~A, Wang, J,  \& Wolynes, P.~G.
\newblock (1997) {\em Proc. Nat. Acad. Sci. USA} {\bf 94}, 777--782.

\bibitem{PortmanJprl98}
Portman, J.~J, Takada, S,  \& Wolynes, P.~G.
\newblock (1998) {\em Phys Rev Lett} {\bf 81}, 5237--5240.

\bibitem{Klimov98}
Klimov, D \& Thirumalai, D.
\newblock (1998) {\em J Mol Biol} {\bf 282}, 471--492.

\bibitem{Brookspnas98}
Sheinerman, F.~B \& Brooks, C.~L.
\newblock (1998) {\em Proc Nat Acad Sci USA} {\bf 95}, 1562--1567.

\bibitem{Micheletti99}
Micheletti, C, Banavar, J.~R, Maritan, A,  \& Seno, F.
\newblock (1999) {\em Phys Rev Lett} {\bf 82}, 3372--3375.

\bibitem{SheaJE99}
Shea, J.~E, Onuchic, J.~N,  \& Brooks, C.~L.
\newblock (1999) {\em Proc Nat Acad Sci USA} {\bf 96}, 12512--12517.

\bibitem{NymeyerH00:pnas}
Nymeyer, H, Socci, N.~D,  \& Onuchic, J.~N.
\newblock (2000) {\em Proc Nat Acad Sci USA} {\bf 97}, 634--639.

\bibitem{SocciND95:jcp}
Socci, N.~D \& Onuchic, J.~N.
\newblock (1995) {\em J Chem Phys} {\bf 103}, 4732--4744.

\bibitem{AbkevichVI95}
Abkevich, V.~I, Gutin, A.~M,  \& Shakhnovich, E.~I.
\newblock (1995) {\em J Mol Biol} {\bf 252}, 460--471.

\bibitem{BetancourtMR95}
Betancourt, M.~R \& Onuchic, J.~N.
\newblock (1995) {\em J Chem Phys} {\bf 103}, 773--787.

\bibitem{VigueraAR96}
Viguera, A.~R, Villegas, V, Aviles, F.~X,  \& Serrano, L.
\newblock (1996) {\em Folding and Design} {\bf 2}, 23--33.

\bibitem{Plaxco98}
Plaxco, K.~W, Simons, K.~T,  \& Baker, D.
\newblock (1998) {\em J Mol Biol} {\bf 277}, 985--994.

\bibitem{ShoemakerWang99}
Shoemaker, B.~A, Wang, J,  \& Wolynes, P.~G.
\newblock (1999) {\em J Mol Biol} {\bf 287}, 675--694.

\bibitem{AlmE99}
Alm, E \& Baker, D.
\newblock (1999) {\em Proc Nat Acad Sci USA} {\bf 96}, 11305--11310.

\bibitem{MunozV99}
Munoz, V \& Eaton, W.~A.
\newblock (1999) {\em Proc Nat Acad Sci USA} {\bf 96}, 11311--11316.

\bibitem{AndersonPW92}
Anderson, P.~W.
\newblock (1992) {\em Concepts in solids}.
\newblock (Addison-Wesley, Reading, Massachusets).

\bibitem{PercusJK82}
Percus, J.~K.
\newblock (1982) in {\em The liquid state of matter: Fluids, simple and
  complex}, eds.{} Montroll, E \& Lebowitz, J.
\newblock (North-Holland, Amsterdam).

\bibitem{PlotkinSS96}
Plotkin, S.~S, Wang, J,  \& Wolynes, P.~G.
\newblock (1996) {\em Phys Rev E} {\bf 53}, 6271--6296.

\bibitem{FloryPJ56:jacs}
Flory, P.~J.
\newblock (1956) {\em J Am Chem Soc} {\bf 78}, 5222--5235.

\bibitem{DillKA93:pnas}
Dill, K.~A, Fiebig, K.~M,  \& Chan, H.~S.
\newblock (1993) {\em Proc Nat Acad Sci USA} {\bf 90}, 1942--1946.

\bibitem{Munoz96:Rev}
Munoz, V \& Serrano, L.
\newblock (1996) {\em Folding and Design} {\bf 1}, R71--R77.

\bibitem{HagenSJ96:pnas}
Hagen, S.~J, Hofrichter, J.~A, Szabo, A,  \& Eaton, W.~A.
\newblock (1996) {\em Proc Nat Acad Sci USA} {\bf 93}, 11615--11617.

\bibitem{KimDE98}
Kim, D.~E, Gu, H,  \& Baker, D.
\newblock (1998) {\em Proc Nat Acad Sci USA} {\bf 95}, 4982--4986.

\bibitem{BrownBM99}
Brown, B.~M \& Sauer, R.~T.
\newblock (1999) {\em Proc Nat Acad Sci USA} {\bf 96}, 1983--1988.

\bibitem{Oxtoby96}
Karpov, V.~G \& Oxtoby, D.~W.
\newblock (1996) {\em Phys Rev B} {\bf 54}, 9734--9745.

\bibitem{KolinskiA93:jcp}
Kolinski, A, Godzik, A,  \& Skolnick, J.
\newblock (1993) {\em J Chem Phys} {\bf 98}, 7420--7433.

\bibitem{Wolynes97nsb}
Wolynes, P.~G.
\newblock (1997) {\em Nature Struct Biol} {\bf 4}, 871--874.

\end{thebibliography}

\end{document}